\documentclass[nofootinbib,twocolumn,pra,letterpaper,longbibliography,superscriptaddress]{revtex4-2}
\usepackage{latexsym,amsmath,amssymb,amsfonts,graphicx,color,amsthm}
\newtheorem{theorem}{Theorem}
\usepackage{enumerate}
\usepackage{braket}
\usepackage{adjustbox}
\usepackage{footnote}
\usepackage[dvipsnames]{xcolor}
\usepackage{tipa}
\usepackage{textcomp}
\usepackage{fontenc}
\usepackage{mathrsfs}
\usepackage{color}
\usepackage{colortbl}
\usepackage{graphics,graphicx}
\usepackage{txfonts}
\usepackage{caption}
\usepackage{subcaption}

\usepackage{amsmath}
\usepackage{amsfonts,amssymb,amstext,amscd,amsthm,bbold}
\usepackage{comment,float}
\usepackage[colorlinks=true,linkcolor=blue,urlcolor=magenta,citecolor=magenta]{hyperref}

\definecolor{myurlcolor}{rgb}{0,0,0.7}

\usepackage{float}
\usepackage{hyperref}

\newcommand{\pt}{\mathcal{P}\mathcal{T}}        
\newcommand{\G}{\mathcal{G}}
\renewcommand{\ket}[1]{\vert#1\rangle}

\renewcommand{\bra}[1]{\langle#1\vert}
\newcommand{\miniprod}[2]{\langle#1\vert#2\rangle}

\begin{document}
\title{Non-Hermitian quantum walks and non-Markovianity: the coin-position interaction} 
\author{Himanshu Badhani}
\email{himanshub@imsc.res.in}
\affiliation{The Institute of Mathematical Sciences, C. I. T. Campus, Taramani, Chennai 600113, India}
\affiliation{Homi Bhabha National Institute, Training School Complex, Anushakti Nagar, Mumbai 400094, India}
\author{Subhashish Banerjee}
\affiliation{Indian Institute of Technology Jodhpur, Jodhpur-342011, India}
\author{C. M. Chandrashekar}
\email{chandru@imsc.res.in}
\affiliation{The Institute of Mathematical Sciences, C. I. T. Campus, Taramani, Chennai 600113, India}
\affiliation{Homi Bhabha National Institute, Training School Complex, Anushakti Nagar, Mumbai 400094, India}
\affiliation{Department of Instrumentation \& Applied Physics, Indian Institute of Science, Bengaluru 560012, India}

\begin{abstract}
A $\pt$-symmetric, non-Hermitian Hamiltonian in the $\pt$-unbroken regime can lead to unitary dynamics under the appropriate choice of the Hilbert space. The Hilbert space is determined by a Hamiltonian-compatible inner product map on the underlying vector space, facilitated by a ``metric operator". A more traditional method, however, involves treating the evolution as open system dynamics, and the state is constructed through normalization at each time step. In this work, we present a comparative study of the two methods of constructing the reduced dynamics of a system evolving under a $\pt$-symmetric Hamiltonian. Our system is a one-dimensional quantum walk with the spin and position degrees of freedom forming its two subsystems. We compare the information flow between the subsystems under the two methods. We find that under the metric formalism, a power law decay of the information backflow to the subsystem gives a clear indication of the transition from $\pt$-unbroken to the broken phase. This is unlike the information backflow under the normalized state method. We also note that even though non-Hermiticity models open system dynamics, pseudo-Hermiticity can increase entanglement between the subsystem in the metric Hilbert space, thus indicating that pseudo-Hermiticity cases can be seen as a resource in quantum mechanics.
\end{abstract}

\maketitle
\section{Introduction}

\begin{figure}
\includegraphics[scale=0.2]{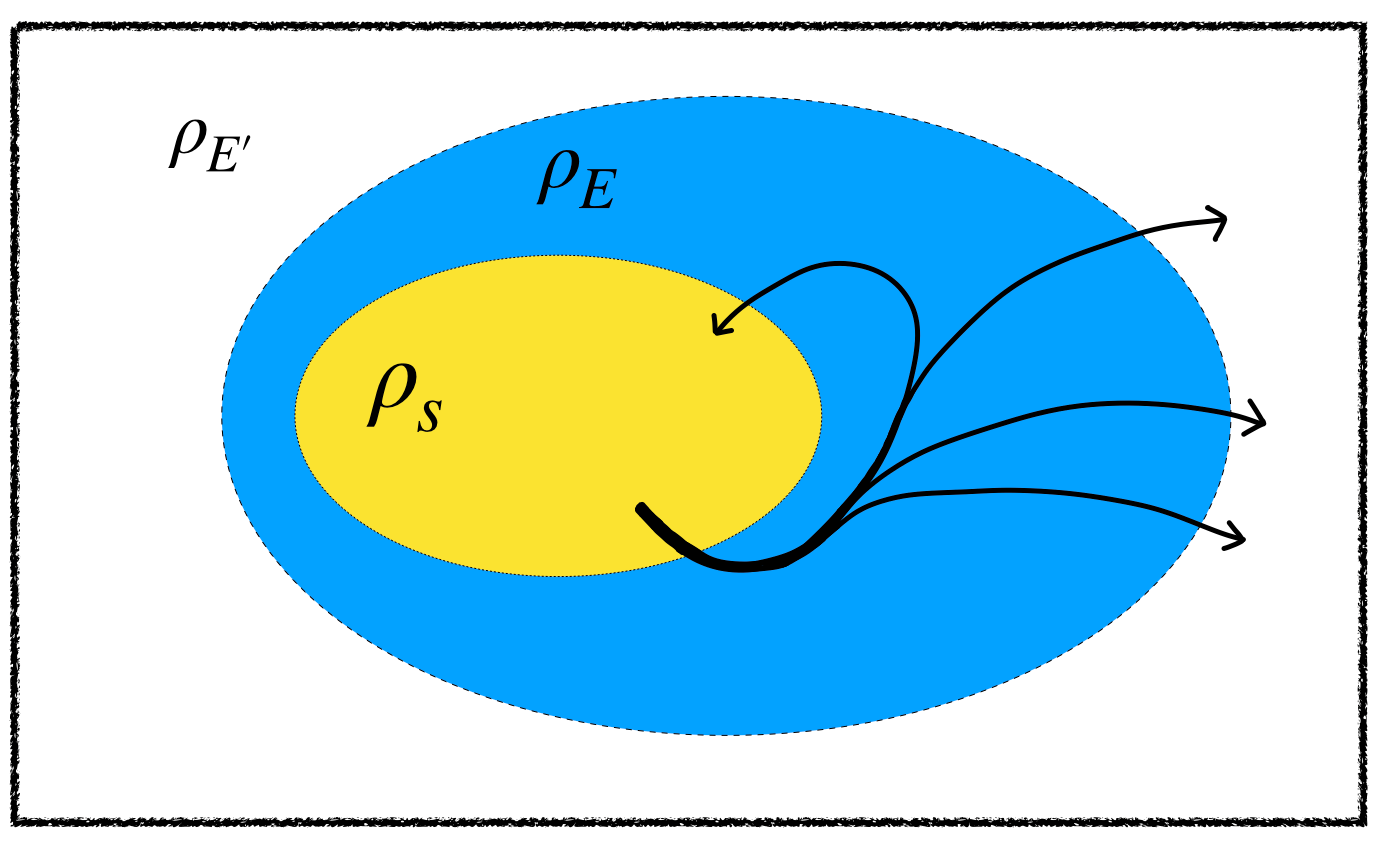}
\caption{Schematic diagram of the problem considered in this work. We aim to study the dependence of the information flow from and to a subsystem when the entire system undergoes a $\pt$-symmetric non-Hermitian evolution.}
\label{opensys}
\end{figure}

The operator formalism of quantum mechanics dictates that corresponding to every physical measurement of a closed system, there is an associated operator (called an observable) whose eigenvalues are the only possible results of the measurement. For this reason, the observables must have only real eigenvalues. 
If the operator is Hermitian, the eigenvalues are guaranteed to be real. However, in 1998 it was shown that there exists a class of non-Hermitian operators that also have real eigenvalues\,\cite{Bender98, Bender02}. These operators are invariant under the Parity-Time reversal (or the $\pt$) operation, i.e., under the consecutive action of time reversal ($t\rightarrow -t$) and spatial reflection ($x\rightarrow -x$). The evolution under $\pt$-symmetric non-Hermitian Hamiltonian is non-unitary, indicating that it models the evolution of an open quantum system \cite{brpet, sbbook}. This has led to a significant interest in the $\pt$-symmetric dynamics in recent years \cite{bender2,rotter,heinrichs,zhang,kishore,anirban,lg}, including the introduction of the so called pseudo-Hermiticity\,\cite{Ali03} which proposes an equivalence between the unitary and $\pt$-symmetric non-unitary dynamics with a redefinition of the inner product. Quantum walks, a unitary evolution and a quantum analog of the classical random walks hold a lot of promise for quantum algorithms and quantum simulations \cite{QWChandru1,SymNoise,QWRel,QWParrondo}. The $\pt$-symmetric non-unitary version of quantum walks was first defined in 2016\,\cite{Ken16}. In this work, we will try to characterize the interaction between the position and coin spaces of a one-dimensional discrete-time quantum walk under non-Hermitian (non-unitary) evolution. The reduced dynamics of a system, obtained by tracing over the ambient environment can be conveniently treated within the framework of Open Quantum Systems. This dynamic is  generally non-Markovian  \cite{rhp,vega,hall,sss,NMdehasing,NMQW,NMdisorder,NNMQCorls}, and one of the major mechanisms responsible for it is the information back-flow from the environment to the system \cite{blprev}. This is also observed in the dynamics of the internal coin state of the walk\, \citep{Jav20}. This non-Markovianity can play an important role in quantum algorithms and quantum communications \cite{NMPingPong} that utilize the properties of quantum walks. It thus becomes important to understand how the reduced dynamics, especially its non-Markovian characteristics, get affected due to the non-Hermiticity of the total quantum walk. In this work, we study a one-parameter family of walk operators generated by $\pt$-symmetric non-Hemritian Hamiltonian and study the divisibility of the reduced (coin) dynamics and the entanglement between the position and coin spaces.
\\
Figure \ref{opensys} shows a schematic representation of the scenario studied in this work. A system with two different degrees of freedom, such as the spin and position of a quantum walk, interacts with the environment under the $\pt$-symmetric non-Hermitian Hamiltonian. The coin space of the 1-dimensional quantum walk plays the role of the subsystem of interest ($\rho_S$), the position space is the complementary subsystem, or the immediate environment ($\rho_E$) and the whole system evolves under a $\pt$-symmetric non-Hermitian Hamiltonian. Our aim in this work is to see the effect of non-Hermiticity on the interaction between the two subsystems $\rho_S$ and $\rho_E$. With the introduction of non-Hermiticity, we effectively increase the degrees of freedom of the environment with which our subsystem, i.e.  the coin space, is interacting. One therefore expects the information back-flow to the subsystem to decrease with the introduction of non-Hermiticity. The back-flow however depends on the dynamics of the system as a whole, including the subsystem interactions. In this study, we will mainly investigate this question using the two primary methods of dealing with the non-Hermitian evolution.  The information back-flow to the coin space will therefore depend on the walk parameters and the parameters controlling the non-Hermitian part of the Hamiltonian.
This article is structured as follows: we introduce the $\pt$-symmetric non-Hermiticity and its treatment under the metric formalism operator in section \ref{sec2}. In section \ref{sec3} we introduce a $\pt$-symmetric non-Hermitian quantum walk and calculate the metric operator. An important part of this work is the prescription to construct a subsystem under the metric formalism when the system can be described as a section over a direct sum of Hilbert spaces. In section \ref{sec4} we analyse the information flow in the reduced coin space in the above two formalisms. We also study the coin-position entanglement under non-Hermiticity in \ref{sec4-2}. We conclude with a short discussion of our results in section \ref{sec5}.
\section{Non-Hermitian quantum mechanics}\label{sec2}
The Hamiltonian of the system governs the time evolution of a system. In quantum mechanics, the Schrödinger equation gives us the governing dynamics of a quantum state $\psi$ in terms of a Hamiltonian operator,
\begin{equation}\label{schr}
	i\partial_t\ket{\psi}=H\ket{\psi}.
\end{equation}
The corresponding equation of motion for the density matrix is given by 
\begin{equation}\label{nonhermeq}
	i\partial_t\rho=H\rho-\rho H^\dagger.
\end{equation}
which reduces to the von Neumann equation for the case of Hermitian Hamiltonian: $i\partial_t\rho=[H,\rho]$.  A density matrix $\rho$ can be interpreted as a mixture of pure states: $\rho=\sum_i p_i\ket{\psi_i}\bra{\psi_i}$ where the state $\ket{\psi_i}$ is prepared with probability $p_i$. This implies that a valid density matrix must have trace $tr(\rho)=\sum_ip_i=1$. It is clear that the trace of a state under unitary transformation remains unchanged: $tr(U\rho U^\dagger)=tr(\rho)$.
\\
Non-Hermitian Hamiltonians appear in many physical situations. A crucial among them is the evolution of an open system under a Lindbladian-like equation \citep{gks,lind}
\begin{equation}\label{lindblad1}
	\partial_t\rho_c(t)=\mathcal{L}_t\rho_c(t)=-i[H,\rho_c(t)]+\sum_j A_j\rho_c(t)A_j^\dagger-\dfrac{1}{2}\{A_jA_j^\dagger,\rho_c(t)\}.
\end{equation}
Here $H$ is assumed to be a Hermitian operator. With a redefinition	$H_e:=H-i\sum_j A_j^\dagger A_j$, this equation can be recast into a form that is a sum of evolution under a non-Hermitian Hamiltonian $H_e$ and an additional term,
\begin{equation}\label{lindbald2}
\mathcal{L}_t\rho_c(t)=-i[H_e\rho_c(t)-\rho_c(t)H_e^\dagger ]+\sum_j A_j\rho_c(t)A_j^\dagger.
\end{equation}
For a two-level system in continuous observation, its dynamics can be cast in the above form with the final term $\sum_j A_j\rho_c(t)A_j^\dagger$ contributing to the jump of the state from the higher energy state to the lower energy state. It is therefore called a called the ``jump term". During such continuous observations, post-selecting the trajectories where one does not see a jump till a time $t$, the dynamics can be given purely in terms of a non-Hermitian Hamiltonian.
\\ \\
Non-Hermitian Hamiltonians also arise in the optical settings\, \citep{nat1} where the refractive index can be adjusted in a manner such that the losses in one direction are matched by the gains in another direction. In terms of the Hamiltonian in the Schrödinger equation, this is equivalent to the potential being symmetric under the simultaneous action of a linear $\mathcal{P}$-transformation, and an anti-linear $\mathcal{T}$-transformation. Such so-called $\pt$-symmetric operators are known to have eigenvalues that come in conjugate pairs and in certain cases have a real spectrum (the so-called unbroken $\pt$-symmetry). In this work, we will focus our attention on such Hamiltonians.
\\
For non-Hermitian Hamiltonians, the evolution operator is not unitary, which renders the trace of the density matrix $\rho$ time-dependent. One way to get a trace 1 density matrix is to redefine it as 
\begin{equation}\label{normalised}
\rho\rightarrow\rho_N=\rho/tr(\rho).
\end{equation} 
We call this the trace normalization method. This redefined density matrix satisfies the equation of motion
\begin{equation} 	
\begin{aligned}
i\partial_t\rho_N&=H\rho_N-\rho_N H^\dagger-\rho_N\dfrac{\partial_ttr(\rho)}{tr(\rho)}\\
&= H\rho_N-\rho_N H^\dagger-i\rho_Ntr\{\rho_N,H^\dagger-H\}.
\end{aligned}
\end{equation}
Such a non-linear evolution of the density matrix has been shown to disobey the no-signaling theorem \citep{Lee14, Brody16}. 
\\\\
\textbf{Metric formalism:} The normalization method of constructing the density matrix is akin to adjusting the probability of the complete set of possibilities to 1. However, this method renders the dynamical equation of the state to be non-linear. For certain non-Hermitian evolutions, the dynamics can be shown to be not just linear, but even unitary. This happens when one considers the pseudo-Hermitian Hamiltonians which satisfy the following relation for a Hermitian \textit{metric} operator $G$\, \citep{Ali10,Ali02}:
\begin{equation}
H^\dagger G=HG.
\end{equation}
The operator $G$ is called a metric operator since one can define an inner product $\miniprod{\psi}{\phi}_G\equiv\miniprod{\psi}{G\phi}$. This altered inner product defines the Hilbert space $\mathcal{H}_G$ which we will call a non-trivial metric space as opposed to a trivial, $G=\mathbb1$ metric Hilbert space (or the Euclidean metric space). In such Hilbert spaces, any operator that satisfies the pseudo-Hermiticity property has real expectation values: $\miniprod{\psi}{H\psi}_G=\miniprod{\psi}{GH\psi}=\miniprod{\psi}{H^\dagger G \psi}$. The adjoint of an operator $\mathcal{O}$ in this Hilbert space is given by $\mathcal{O}^\#=G^{-1}\mathcal{O}^\dagger G$ as one can show :
\begin{equation}
\begin{aligned}
\text{by definition:}\hspace{1cm} \miniprod{\mathcal{O}^\#\phi}{\psi}_G&= \miniprod{\phi}{\mathcal{O}\psi}_G\\
\Rightarrow\miniprod{\mathcal{O}^\#\phi}{G\psi}&= \miniprod{\phi}{G\mathcal{O}\psi}\\
&=\miniprod{\mathcal{O}^\dagger G\phi}{\psi}\\
&=\miniprod{G^{-1}\mathcal{O}^\dagger G\phi}{G\psi}
\end{aligned}
\end{equation}
The $\mathcal{O}^\#$ operation is sometimes called the \textit{generalized dagger operation} and the pseudo-Hermiticity condition can be expressed as the self-adjointness condition: $\mathcal{O}^\#=\mathcal{O}$. Consequently, any operator that can be written as exponential of a pseudo-Hermitian operator, i.e. $U=e^{-iH}$ plays the role of a unitary operation in the sense that it preserves the inner product: $\miniprod{U\psi}{U\psi}_G=\miniprod{\psi}{\psi}_G$ since $U^\#=U^{-1}$. A density matrix in this space must also be pseudo-Hermitian denoted by $\overline{\rho}$ which evolves under the operator $U=e^{iHt}$ as $\overline{\rho}(t)=U\overline{\rho}U^\#$. The equation of motion for such a state is given by the usual von Neumann equation
\begin{equation}
i\hbar\partial_t\overline{\rho}=[H,\overline{\rho}].
\end{equation}
One can go beyond the pseudo-hermiticity regime to define a unitary evolution for the state. Since under unitary evolution, the inner product does not change with time, such a constraint put on the inner product $\miniprod{\phi(t)}{G\psi(t)}$, where $\ket{\psi(t)}=e^{-iHt}\ket{\psi}$, gives the following necessary relation :
\begin{equation}\label{geom}
	i[G(t)H-H^\dagger G(t)]=\partial_tG(t).
\end{equation}
Therefore, for non-pseudo-Hermitian Hamiltonians, the metric operator must be time-dependent. This is equivalent to updating the Hilbert space at each time instance, not unlike the normalization at every time step. However, using the metric makes this evolution a linear evolution. 
The metric operator can be constructed using the right eigenvectors $\ket{n(t)}$ of the Hamiltonian $H(t)$ (\citep{Ali02}), such that
\begin{equation}\label{metric}
	G(t)=\Big(\sum \ket{n(t)}\bra{n(t)}\Big)^{-1},
\end{equation}
where $\ket{n(t)}=e^{-i\epsilon_nt}\ket{n(0)}$ for the right eigenvectors $\ket{n(0)}$ of the Hamiltonian with eigenvalue $\epsilon_n$. From the definition, $G(t)=(\sum_n \ket{n(t)}\bra{n(t)})^{-1}=(\sum_n e^{2\Im(\epsilon_n)t} \ket{n(0)}\bra{n(0)})^{-1}$, where $\Im(\epsilon_n)$ is the imaginary part of $\epsilon_n$. Therefore we have $H G^{-1}=\sum_n \epsilon_n\ket{n(t)}\bra{n(t)}$ and  $G^{-1}H^\dagger=\sum_n \epsilon_n^*\ket{n(t)}\bra{n(t)}$. Using this one can show that the operator in equation \eqref{metric} satisfies equation \eqref{geom}:
\begin{equation} 
\begin{aligned}
 RHS=&i[G(t)H-H^\dagger G(t)]\\
		=& iG(t)[HG(t)^{-1}-G(t)^{-1}H^\dagger]G(t)\\
		=& iG(t)\sum_n (\epsilon_n-\epsilon_n^*)\ket{n(t)}\bra{n(t)}G(t)\\
		=&G(t) \sum_n -2\Im(\epsilon_n)e^{2\Im(\epsilon_n)t} \ket{n(0)}\bra{n(0)}G(t)\\
		=&-G(t)\partial_tG^{-1}(t)G(t)\\
		=&\partial_tG(t)=  LHS.
\end{aligned}
\end{equation}
\noindent 
Finally, one can always choose the metric such that the trace of the density matrix is always 1: $G(t)\rightarrow G(t)/(tr(\rho))$.
\section{Metric formalism for the quantum walk}\label{sec3}
\begin{figure}
\begin{subfigure}{.5\textwidth}
  \includegraphics[scale=.35]{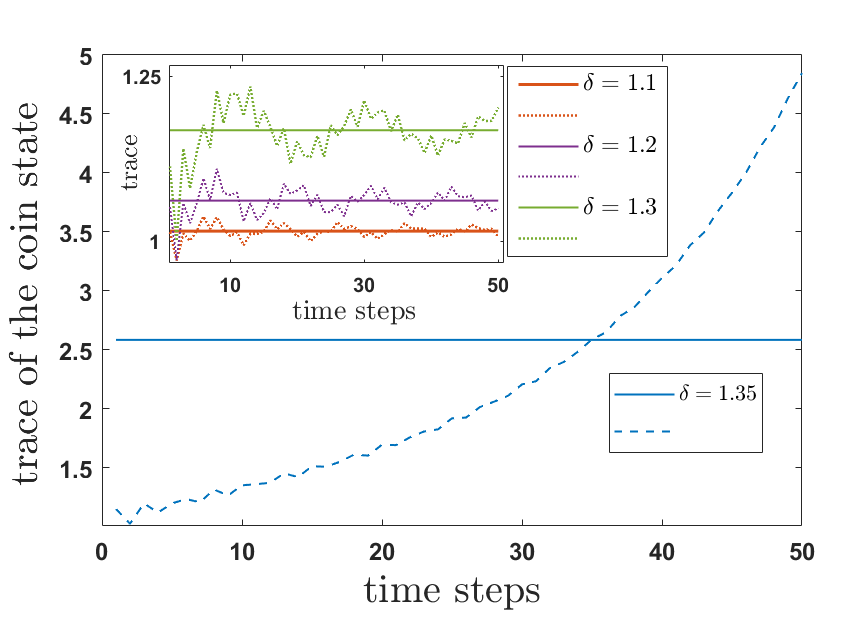}
  \label{trace2}
\end{subfigure}
\caption{Trace of the coin state as a function of time, evolved under a walking operator with coin parameters $\theta_1=\pi/4$ and $\theta_2=-\pi/7$ and the exceptional point is given by $\delta_{PT}\sim1.34714..$ Dotted lines correspond to the non-Hermitian evolution without the metric and solid lines correspond to the state with metric correction given by equation \eqref{momwalk}. Inset corresponds to $\delta$ values below the exceptional point. As expected, the state with metric redefinition has a time-invariant trace.}
\label{tracecomparisions}
\end{figure}

A one-dimensional discrete-time quantum walk is an evolution of a system with two degrees of freedom, called the position and coin, or the external and internal degree of freedom respectively. One step of a quantum walk is typically given by an evolution of the internal state (the coin operation $C$) followed by an evolution in the external state (the shift operation $S$) conditioned on the internal state: $W=SC$. The shift and the coin state for a unitary walk for a spin half particle is given by 
\begin{equation}\label{shift}
\begin{aligned}
	S&=\sum_x\ket{x+1}\bra{x}\otimes\ket{\uparrow}\bra{\uparrow}+\ket{x-1}\bra{x}\otimes\ket{\downarrow}\bra{\downarrow}, \\
	\overline{C}(\theta)&=I_N\otimes\begin{pmatrix}
\cos{\theta} & i\sin{\theta}\\
i\sin{\theta} & \cos{\theta}
\end{pmatrix}=I_N\otimes C(\theta).
\end{aligned}.
\end{equation}
Here $\ket{\uparrow}$ and $\ket{\downarrow}$ are the eigenstates of the $\sigma_3$ operator, and $\theta$ is a parameter of the coin operator. Inspired by the balanced gain-loss models in optics\, \citep{nat1}, a $\pt$-symmetric non-Hermitian version of a quantum walk was introduced by introducing a non-unitary operator in the walk\,\citep{Ken16}. The gain and loss in the amplitudes of the $\ket{\uparrow}$ and $\ket{\downarrow}$ states are executed by the following operators 
\begin{equation}
	\G_2=\G_1^{-1}=\begin{pmatrix}
\delta & 0\\
0 & 1/\delta
\end{pmatrix}
\end{equation}
where $\delta\in\mathbb{R}^+$ parametrizes the degree of non-Hermiticity. The $\pt$-symmetric quantum walk is  then given by the following general walk operator:
\begin{equation}
	W_{nh}=\overline{C}(\theta_1/2)\ S\ \overline{\G}_2\ \overline{C}(\theta_2)\ S\ \overline{\G}_1\ \overline{C}(\theta_1/2),
\end{equation}
where $\overline{\G}=I_N\otimes \G$. For the Hamiltonian that commutes with $\pt$ operation, the operation $W=e^{-iH}$ satisfies:
\begin{equation}\label{ptwalk}
	(\pt) W (\pt)^{-1}=W^{-1}.
\end{equation}
We see that the above given quantum walk operator $W_{nh}$ satisfies this relation and therefore simulates the evolution under a $\pt$-symmetric non-Hermitian Hamiltonian. \\
Let us assume that the quantum walk takes place on a circular lattice. In this case, the shift operator given in\, \eqref{shift} can be written as a direct sum over the Fourier coordinates $k$, 
\begin{equation}
    \begin{aligned}
	S=&\sum_{k}\ket{k}\bra{k}\otimes \begin{pmatrix}
e^{ik} & 0\\
0 & e^{-ik}
\end{pmatrix}\\
=& \sum_{k}\ket{k}\bra{k}\otimes S(k).
\end{aligned}
\end{equation}
Therefore, the walk operator in the Fourier coordinates diagonalizes as 
\begin{equation}\label{ptHamil}
\begin{aligned}
	W_{nh}=\sum_k \ket{k}\bra{k}\otimes W_c(k).
	\end{aligned}
\end{equation}
Here, $W_c(k)$ is a $\pt$-symmetric evolution operator given by
\begin{equation}
\begin{aligned}	
W_c(k)&=S(k) \G^{-1}C(\theta_2)S(k)\G C(\theta_1)\\
\end{aligned}.
\end{equation}
In the case of a discrete-time quantum walk, if we start with an initial state $\overline{\rho}_0=\dfrac{1}{N}\sum_k \ket{k}\bra{k}\otimes \overline{\rho}_k(t=0)$ in the Fourier space representation, at all $t$, the state has the form $\overline{\rho}=\sum_k\ket{k}\bra{k}\otimes \overline{\rho}_k(t)$, where $\overline{\rho}_k(t)=W_c(k)^t\ \overline{\rho}_k(t=0){W_c(k)^\dagger}^t$. 
\\\\
\textbf{Exceptional point:} 
Eigenvalues of the Hamiltonian can be witnessed through the eigenvalues of the walk operator via the relation $W= e^{-iH}$. The eigenvalues of the operator $W_c(k)$ in the Fourier space can be given as,
\begin{equation}\label{eig}
\begin{aligned}
	\lambda_{\pm}&=(a\pm\sqrt{a^2-1}) ,\\
	\text{where, }\\
	\hspace{.25 cm}a&=\cos(2k)\cos(\theta_1)\cos(\theta_2)-\cosh(2\ln(\delta))\sin(\theta_1)\sin(\theta_2).
	\end{aligned}
 \end{equation}
The eigenvalues of the corresponding Hamiltonian will be given by $\epsilon_\pm=i\ln(\lambda_\pm)$. We can solve this equation by a simple substitution: $a=\cos(x)$ after which the walk operator's eigenvalues become $\lambda_\pm=\cos(x)\pm i\sin(x)$. Therefore, $\epsilon_\pm=\mp \cos^{-1}(a)$. The unbroken $\pt$-symmetry is therefore satisfied by the Hamiltonian for which $a\in \mathbb{R}$ and $a<1$. For $\epsilon_{\pm}\in\mathbb{R}$, the eigenvectors are also the eigenvectors of the $\pt$ operator. Therefore $|a|<1$ is the unbroken $\pt$-symmetry regime.
\\
The eigenvalues become degenerate at $a=1$, corresponding to $\epsilon_{\pm}=0$. This is called the exceptional point of the $\pt$-symmetry dynamics and indicates the transition from unbroken $\pt$-symmetry to broken $\pt$-symmetry. Beyond this point, the eigenvalues shift from being real to complex conjugate pairs. For $a>1$,  $\cos^{-1}(a)$ is purely imaginary, so the eigenvalues go from being completely real to completely imaginary. From equation \eqref{eig} it is clear that to satisfy $a>1$, $\theta_1$ and $\theta_2$ must be of the opposite sign as the first term is always smaller than 1. Further, for a given $\delta$, $\theta_1$, and $\theta_2$ the minima of $\epsilon(k)$ in the real regime occurs for $k=n\pi$. This gives  us the following relation for the $\pt$-symmetry breaking
\begin{equation}
\begin{aligned}
	\ln(\delta_{\tiny{PT}})=\dfrac{1}{2}\cosh^{-1}\Big[\dfrac{\cos(\theta_1)\cos(\theta_2)-1}{\sin(\theta_1)\sin(\theta_2)}\Big], \\[.3cm]
	\text{where,}\hspace{1cm} \theta_1\in [0,\pi),\theta_2\in (-\pi,0].
\end{aligned}
\end{equation}
The metric operator is not well defined at the exceptional point and beyond this point it becomes time-dependent as the Hamiltonian is no more pseudo-Hermitian.
\\ 
To construct the metric we first find the eigenvectors of the Hamiltonian which are given by $\{\ket{k}\ket{\phi_i(k)}\}$, $i=\{1,2\}$ where $\ket{\phi_i(k)}$ are the eigenvectors of the operator $W_c(k)$. If the eigenvalues of the corresponding Hamiltonian $H=-i\ln(W_c(k))$ are given by $\epsilon_i(k)$, these eigenvectors will therefore evolve with time as $\ket{\phi_i(k,t)}=e^{-i\epsilon_i(k)t}\ket{\phi_i(k)}$.
\begin{equation}
\begin{aligned}
	G(t)=&\Big(\sum_{k,i}\ket{k}\ket{\phi_i(k,t)}\bra{k}\bra{\phi_i(k,t)}\Big)^{-1}\\
	=&\sum_{k}\ket{k}\bra{k}\otimes\Big(\sum_i \ket{\phi_i(k,t)}\bra{\phi_i(k,t)}\Big)^{-1}\\
	\equiv &\sum_{k}\ket{k}\bra{k}\otimes G_c(k,t).
\end{aligned}
\end{equation}
Therefore the metric operator itself is block-diagonalized in the Fourier space for the isotropic walk operator with the momentum-dependent metric in the coin space given by:
\begin{equation}
	G_c(k,t)=\Big(\sum_{i=1,2}e^{2\Im(\epsilon_i(k))t}\ket{\phi_i(k)}\bra{\phi_i(k)}\Big)^{-1}.
\end{equation}
where $\Im(\epsilon_i(k))$ is the imaginary part of the eigenvalues. The state in the metric space is therefore given by the 
\begin{equation}\label{momwalk}
\begin{aligned}
	\overline{\rho}(t)
	= &\dfrac{1}{N}\sum_{-\pi}^{\pi-dk}\ket{k}\bra{k} \otimes W_c(k)^t\overline{\rho}_c(t=0)(W_c(k)^\#)^t\\
	=&\dfrac{1}{N}\sum_{-\pi}^{\pi-dk} \ket{k}\bra{k}\otimes \overline{\rho}_k(t).
\end{aligned}
\end{equation}
This is a state in the direct sum of Hilbert spaces $\oplus\mathcal{H}_k$ where each Hilbert space $\mathcal{H}_k$ is the vector space $\mathbb{C}^2$ endowed with the metric $G_c(k,t)$. It is clear that if all these metric operators were the same, the reduced state would simply be given by the sum of all the states in these Hilbert spaces: $\overline{\rho}_c(t)=\dfrac{1}{N}\sum_k\overline{\rho}_k(t).$ However, this summation has no meaning if each state $\overline{\rho}_k$ is defined in different metric spaces. This is akin to the problem of adding two vectors at different locations on a curved (non-Euclidean) space. To add operators in two different metric spaces one has to introduce an isomorphic map between the two spaces which preserves the Hermiticity and the norm of the operator from one metric space to the other. In the following theorem, we give such a map :
\\ \\
\begin{theorem}
 Given two Hilbert spaces $\mathcal{H}_{G_k}$ and $\mathcal{H}_{G_l}$ with metric operators $G_k$ and $G_l$. Let $\eta_k$ and $\eta_l$ be the positive square root of these metric operators respectively. There is an isomorphism that maps operators $\mathcal{O}_l\in\mathcal{H}_{G_l}$ to operators in $\mathcal{O}_{k,l}\in\mathcal{H}_{G_k}$ which preserves the pseudo-Hermiticity and the expectation values:
\begin{equation}\label{partrans}
\mathcal{O}_{k,l}=\eta_k^{-1}\ \eta_l\ \mathcal{O}_l\ \eta_l^{-1}\ \eta_k
\end{equation}   
\end{theorem} 
\begin{proof}
The map $\psi\rightarrow \eta\psi$ is an isometry between the Hilbert spaces $\mathcal{H}_{G_l}$ and $\mathcal{H}$ since $\miniprod{\psi}{\psi}_G=\miniprod{\eta\psi}{\eta\psi}$. This implies that $\mathcal{O}\rightarrow \eta_l\mathcal{O}_l\eta_l^{-1}\equiv\mathcal{O}_{\eta_l}$ is a transformation between the space of bounded operators $\mathcal{B}(\mathcal{H}_G)\rightarrow\mathcal{B}(\mathcal{H})$ that preserves the norm $||\mathcal{O}||_{G_l}=||\mathcal{O}_{\eta_l}||$\, \citep{Ali03}. Furthermore, this transformation, being a similarity transformation, takes pseudo--Hermitian operators to Hermitian operators and keeps that spectrum invariant. Therefore, this is a metric-compatible parallel transport of the operators between spaces $\mathcal{B}(\mathcal{H}_{G_l})\rightarrow\mathcal{B}(\mathcal{H})$.\\
Following this argument, we conclude that the map $\mathcal{O}_l\rightarrow\mathcal{O}_{k,l}=\eta_k^{-1}\eta_l\mathcal{O}_l\eta_l^{-1}\eta_k$ is a spectrum-preserving isometry $\mathcal{B}(\mathcal{H}_{G_l})\rightarrow\mathcal{B}(\mathcal{H}_{G_k})$.
\end{proof}
\begin{figure}[h!]
    \centering
    \includegraphics[width=1\linewidth]{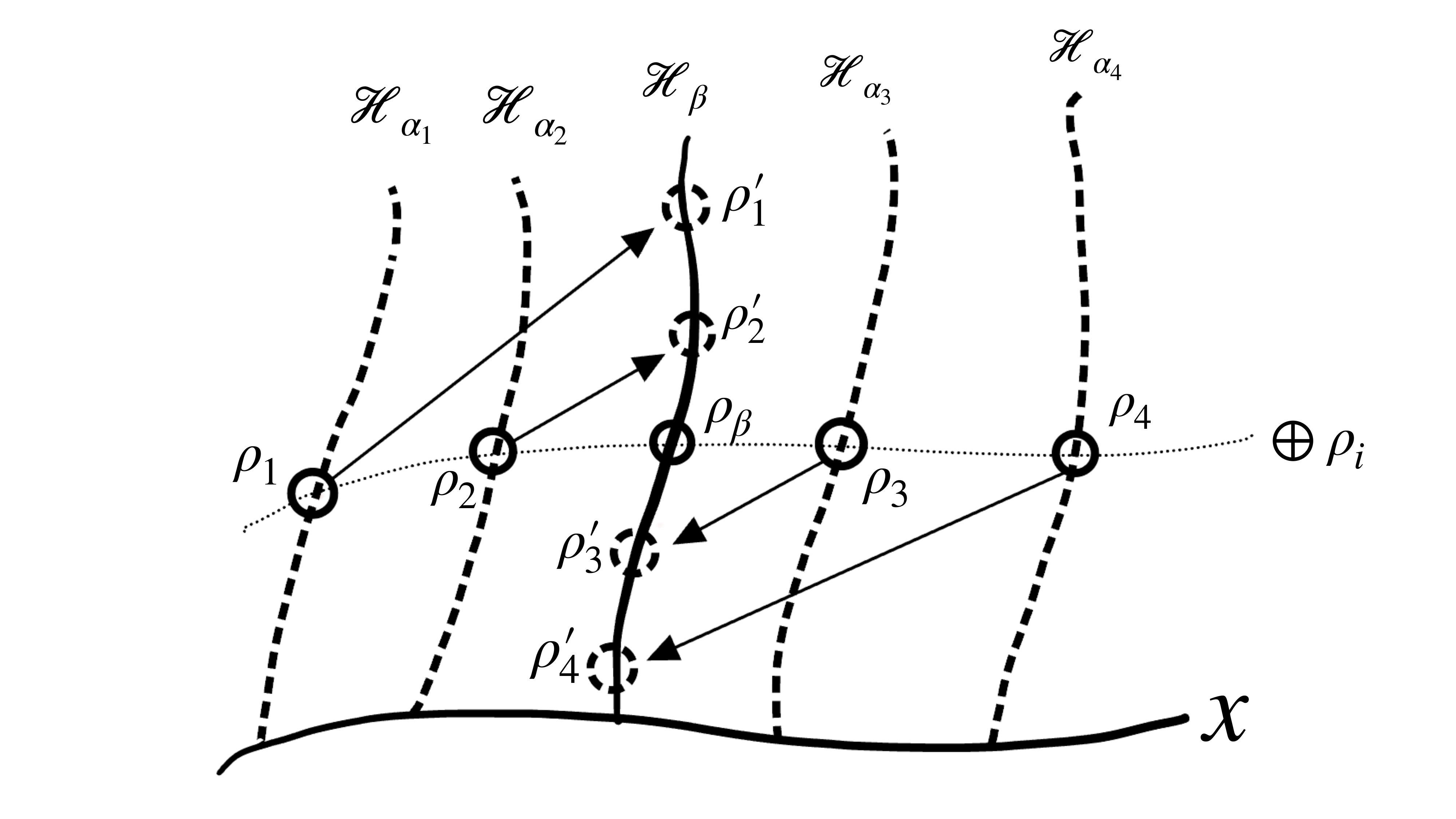}
    \caption{Fiber bundle structure of the Hilbert space, where each fiber is a sub-Hilbert space $\mathcal{H}_{\alpha_i}$ with $x$-dependent metric. If the state has a form of a section over this fiber bundle, $\rho=\oplus \rho_i$, the partial trace operation involves first a parallel transport to any particular fiber $\mathcal{H}_\beta$ via $\rho_i'=\eta_\beta^{-1}\eta_i \rho_i\eta_i^{-1}\eta_\beta$. The reduced state is then given by $\rho_c=\dfrac{1}{N}\sum_i\rho_i'$.}
    \label{fig:fiberbundle}
\end{figure}
\noindent
This also hints at  a non-trivial geometrical structure of the state, similar to the one provided in\, \citep{ali17} (see fig. \ref{fig:fiberbundle}). Let us now construct the reduced coin state of the state $\overline{\rho}$ given in equation \eqref{momwalk}: we transport the states $\overline{\rho}_{k}\in\mathcal{H}_k$ to the states $\overline{\rho}_{k_0,k}\in \mathcal{H}_{k_0}$ per equation \eqref{partrans}, such that we have a unique map between the bounded operators $\mathcal{B}\big(\oplus_k\mathcal{H}_k \big)\rightarrow \mathcal{B}\big(\mathcal{H}_{k_0}^N\big)$. Partial trace operation $\mathcal{B}\big(\mathcal{H}_{k_0}^N\big)\rightarrow \mathcal{B}\big(\mathcal{H}_{k_0}\big)$ is now well defined and gives the following reduced state on $\mathcal{H}_{k_0}$:
\begin{equation}
	\overline{\rho}_c(k_0)=\dfrac{1}{N}\sum_{-\pi}^{\pi-dk} \overline{\rho}_{k_0,k}(t).
\end{equation}
Note that while this reduced state depends on the Hilbert space $\mathcal{H}_{k_0}$, the expectation values $tr(\overline{\rho}_c(k)\overline{\mathcal{O}}_c(k))$ are independent of $k$. Furthermore, is it also clear that this state is equivalent to the virtual subsystem $\rho_\eta^c$ described in \citep{hb23}:
\begin{equation}
\overline{\rho}_c(k_0)=\eta_k^{-1}\rho_\eta^c\eta_k.
\end{equation}
Figure \ref{tracecomparisions} shows the trace of the reduced state constructed under the two formalisms without normalization. The metric method of equation \eqref{momwalk} gives us a time-independent trace and therefore, by changing the metric by a constant factor we can define a trace 1 density matrix. Below the exceptional point (the inset), the trace of the density matrices without metric correction oscillates around its asymptotic values, a trademark for the $\pt$-symmetric evolution. Above the exceptional point (which in the corresponding figure is $\delta_{PT}\sim 1.347$ for the chosen evolution parameters), the trace increases exponentially with time.
\section{Coin-position interaction}\label{sec4}
Our objective in this work is to analyze the interaction between the coin and position degrees of freedom of a 1-dimensional quantum walk. Aspects of this interaction can be studied through the dynamical properties of the coin state. Studying the coin state is experimentally and numerically more feasible since coin degrees of freedom are more easily manipulated and they exist in Hilbert spaces of much smaller dimensions. Here we will study two aspects of the coin-position interaction as the quantum walk as a whole undergoes a a $\pt$-symmetric non-Hermitian evolution: A) the information back-flow to the coin state from the position space and B) the entanglement between the coin and position degrees of freedom. We will study these properties through the dynamics of the coin state which is constructed through the previously discussed methods: the metric formalization and the normalization method.
\\ \\
\subsection{Information back-flow}
\begin{figure}[h!]
 \includegraphics[scale=.38]{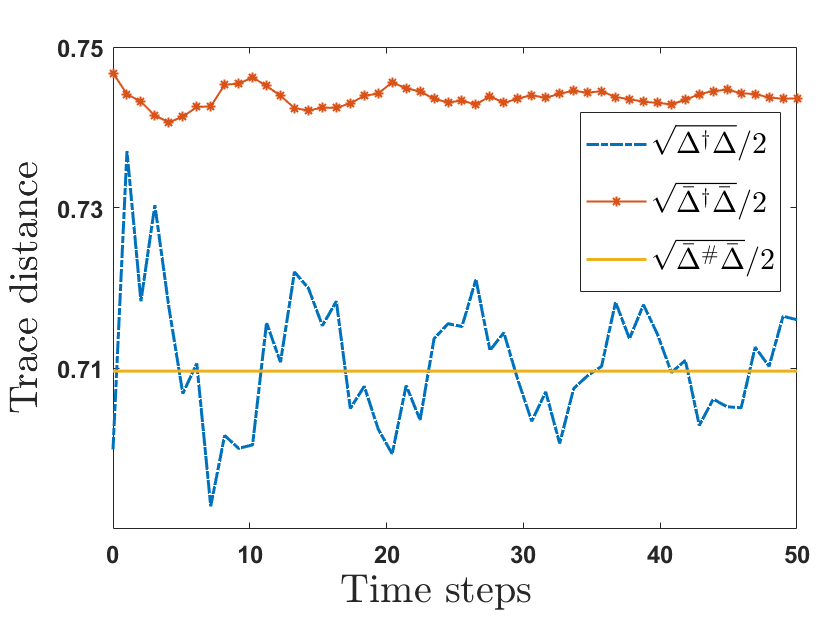}
 \caption{With initial states $\rho(t=0)=\ket{\uparrow}\bra{\uparrow}\otimes\ket{0}\bra{0}$ and $\sigma(t=0)=\ket{\uparrow+\downarrow}\bra{\uparrow+\downarrow}/2\otimes\ket{0}\bra{0}$, the figure shows the trace distance between the two states under the discussed definitions of the trace distance. We have defined $\Delta\equiv\rho-\sigma$ and $\overline{\Delta}\equiv\overline{\rho}-\overline{\sigma}$. The states evolved under the normalized state method are denoted by $\rho$ and $\sigma$, and the ones evolved under the metric formalism are denoted by $\overline{\rho}$ and $\overline{\sigma}$. The non-Hermitian parameter is set to $\delta=1.2$ and the coin parameters are $\theta_1=\pi/4$ and $\theta_2=-\pi/7$ which correspond to the exceptional point $\delta=1.34713.. $. We see that the trace distance is non-contractive under the non-Hermitian, albeit positive trace-preserving, map. In fact, above the exceptional point, the trace distance can be found to be exponentially increasing. However, under the metric formalism the trace distance between the two states is constant.}
 \label{tracedist}
 \end{figure}

\begin{figure}[!h]
\begin{subfigure}{.5\textwidth}
  \centering
  \includegraphics[scale=.33]{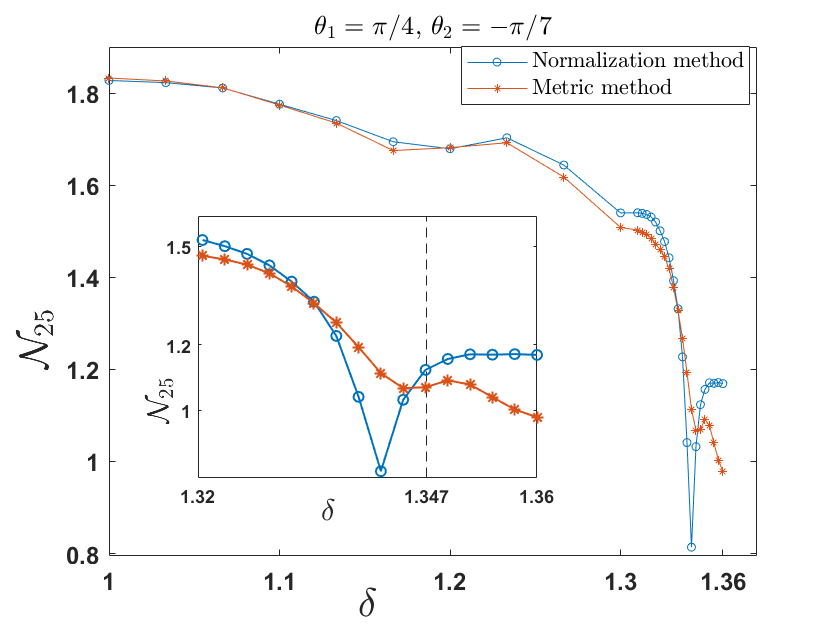} 
  \caption{}
\end{subfigure}
\begin{subfigure}{.5\textwidth}
  \centering
  \includegraphics[scale=.35]{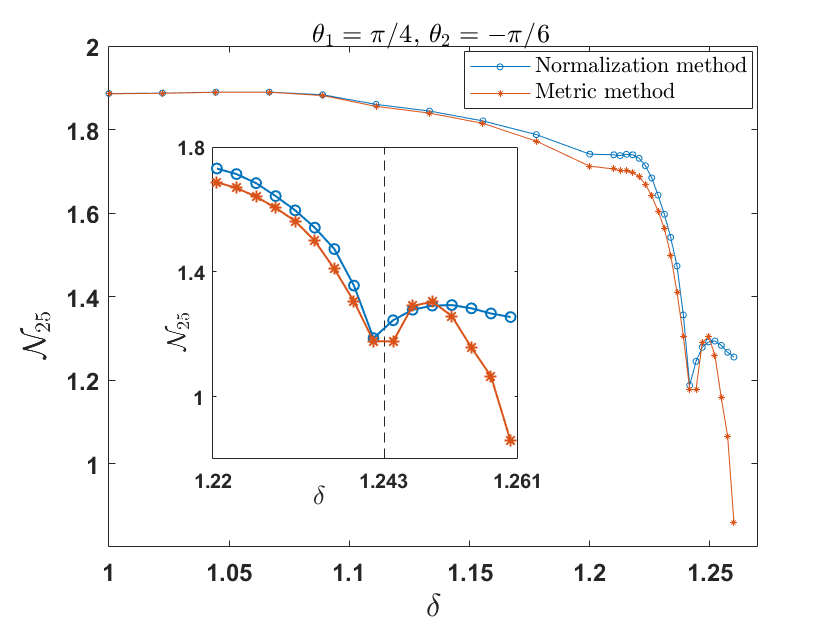} 
  \caption{}
\end{subfigure}
\caption{The measure of information back-flow after 25 steps of the quantum walk, plotted as a function of the dissipation parameter $\delta$ under different quantum walk parameters $\theta_1$ and $\theta_2$. The inset zooms at the region around the exceptional point. Note that the information back-flow in the normalization method does not diverge exactly at the exceptional point. In the main text, we explain that this is a finite time effect.}
\label{blpwithg}
\end{figure}

Whenever a system interacts with its surroundings, partial information about the state of the system is shared with the environment. This information stored in the environment may affect the interaction between the system and the environment at a later time, thus making the evolution of the state depend on its history. This is the mechanism of non-Markovian evolution.
\\ 
Under a stochastic evolution, two states continue losing information about the initial state consequently becoming less and less distinguishable. The distinguishability between two states is directly related to the trace distance between them and, therefore, under a stochastic evolution, characterized by a positive trace-preserving quantum operation (PTP maps), the trace distance between any two states must monotonically decrease \citep{BLP10}. Therefore, a deviation from the monotonic contraction indicates the back-flow of information from the environment to the system.
\\ \\
The trace distance between two states $\rho_1$ and $\rho_2$ is half of the trace norm of $\rho_1-\rho_2$, i.e.,
\begin{equation}
	D(\rho_1,\rho_2)=||\rho_1-\rho_2||_1/2.
\end{equation}
The trace norm of an operator $X$ is defined as $||X||_1=tr(\sqrt{X^\dagger X})$. Under a trace-preserving, positive, linear quantum map $\Lambda$, the trace distance between two states is monotonically decreasing,
\begin{equation}\label{contractivity1}
	D(\Lambda(\rho_1),\Lambda(\rho_2))\le D(\rho_1,\rho_2).
\end{equation}
Equality in the above relation holds only for unitary maps. However, this line of reasoning fails when the evolution is non-linear. For instance, the map in equation \eqref{normalised} is a positive and trace-preserving map, but the trace distance between any two states under this evolution is non-monotonic. Therefore, a violation of monotonicity is not a signature of P-indivisibility under a non-linear evolution and does not characterize the non-Markovianity in the usual sense, see e.g. \citep{Kawa17,Jahromi22}. Nevertheless, the decrease or increase in distinguishability is a reliable indicator of the flow of information from the system or back to the system respectively. We will therefore use the trace distance to measure information flow in the normalised state method. Since the map \eqref{normalised} preserves Hermiticity, the trace distance between two states $\rho_N(t)$ and $\sigma_N(t)$ is simply the sum of the absolute values of the eigenvalues of the operator $\Delta=\rho_N(t)-\sigma_N(t)$.
\\
In the metric picture, the trace norm is given by
\begin{equation}
 	 ||X||_1=tr(\sqrt{X^\# X}).
 \end{equation}
where $X^\#=G^{-1}X^\dagger G$ is the adjoint of the operator $X\in\mathcal{B}(\mathcal{H}_G)$. Under a pseudo-Hermitian Hamiltonian, the trace distance between any two states, constructed under the metric formalism, remains constant since the operator $U=e^{-iH}$ preserves the inner product. Trace distance under each of the above-discussed evolutions is plotted in Fig. \ref{tracedist} where the quantum operation is the $\mathcal{PT}$-symmetric quantum walk operation. We see that with the redefined trace norm the trace distance under the metric formalism is constant whereas the trace distance under the normalized state method oscillates.
Suppose we are given two initial states $\rho$ and $\sigma$ prepared in the Euclidean Hilbert space. At time $t=0$, one introduces a $\pt$-symmetric potential. The corresponding states in the metric space $\mathcal{H}_G$ are given by $\overline{\rho}=\rho G$\, \citep{karvade22}. With this prescription, the evolution $\overline{\rho}(t)=W^t\overline{\rho}(W^{\#})^t$ is a trace-preserving operation. If one uses the normalization method, the states at time $t$ will be denoted by $\rho_N(t)=\dfrac{W^t\rho (W^\dagger)^t}{tr(\rho(t))}$.
\\ \\
Let $D(\Lambda(\rho),\Lambda(\sigma))-D(\rho,\sigma)=\bigtriangledown$ denote the change in the trace distance due to the discrete map $\Lambda$ given either under the state normalization method or the metric formalism.
The measure of information back-flow\, \citep{BLP10}, depending on the initial states $\rho$ and $\sigma$, for these discrete dynamics is given by:
\begin{equation}
\begin{aligned}
N(t,\rho,\sigma)=
\begin{cases}
& N(t-1,\rho,\sigma)+\bigtriangledown, \hspace{.5cm} \text{if $\bigtriangledown>0$}\\
& N(t-1,\rho,\sigma), \hspace{.5cm} \text{if $\bigtriangledown\le 0$}
\end{cases}
\end{aligned}
\end{equation}
where $N(0,\rho,\sigma)=0$. If one maximizes over the set of all initial states $\rho$ and $\sigma$, we get a measure of information back-flow which is a characteristic of the map $\Lambda$ alone:
\begin{equation}
	\mathcal{N}_t:=\max_{\rho,\sigma}\{N(t,\rho,\sigma)\}.
\end{equation}
Figure \ref{blpwithg} shows the information back-flow $\mathcal{N}_t$ after 25 steps of the walk and its variation with the dissipation factor $\delta$.  The maximization of the measure was performed using the Monet Carlo simulated annealing \citep{mc} technique. This technique has the advantage of avoiding local equilibrium points and therefore provides a more reliable optimization. Like every optimization technique, there is an inherent error in the estimation of the BLP measure, which in our simulation can be taken to be of the order $10^{-2}$ units. For the discrete quantum walks, the map $\Lambda$ is given by tracing out the position degrees of freedom from the walk state evolved under either of the above two formalisms. We enumerate the notable observations from this figure:
\begin{enumerate}
\item As one would expect, the information back-flow for both methods is the same for the Hermitian or $\delta=1$ case. At this point the metric becomes trivial and the reduced dynamics is the same for the two methods.
\item The measure starts to deviate for the methods with significantly different behavior near and beyond the exceptional point.
\item The information back-flow to the coin state decreases monotonically (within the error bar of $\pm 10^{-2}$) till near the exceptional point for both methods. 
\item Near the exceptional point, we see a non-monotonicity of information back-flow for both methods. However, detecting non-monotonicity in the normalized state method does not accurately indicate the exceptional point.
\item The difference between the two methods is most stark in the $\pt$ broken regime (beyond the exceptional point). 
\end{enumerate}
Given that the dissipation factor $\delta$ can be seen as a measure of the strength of the interaction between the system and the environment, we can develop some intuitive reasoning on how the information back-flow should vary with $\delta$. As the interaction strength increases, one would assume that, in general, the information of the subsystem would dissipate more easily into the environment, resulting in reduced back-flow to the subsystem. This is exactly what happens in all three figures for both the methods of construction of reduced dynamics. 
\\
The sharp decrease of the information back-flow as we approach the exceptional point is indicative of a phase transition; here the EP plays the role of the critical point near which the information flow shows a power law behavior with respect to the parameter $\delta_{PT}-\delta$ $i.e.$ $\mathcal{N}_t(\delta)\propto (\delta_{PT}-\delta)^p$ as $\delta_{PT}-\delta \rightarrow 0^+$.  The criticality of the information flow between the system and the environment near the exceptional point is a well-known phenomenon \citep{Kawa17,Heiss}, however, it has previously not been studied in regards to the flow of information \textit{within} the system.
\\
Figure \ref{tracedist} shows that within the PT-unbroken regime, there is no information back-flow to the system in the non-trivial metric Hilbert space. Even in Euclidean Hilbert space, the distinguishability shows a periodic behavior and it is known that this period diverges to infinity at the exceptional point\, \citep{Kawa17}. These observations explain the power law behavior of the information back-flow to the subsystem in both methods. The curious case of the minima \textit{before} the EP for the normalization method in figure \ref{blpwithg} is due to the fact that the period of the information retrieval is more than $t=25$ for $\delta\sim 1.343$. For this reason, the normalization method gives errors in the detection of the EP through the decay in information back-flow. A remedy for such errors is to check for information back-flow for a large enough time. The metric method on the other hand has no such periodicity dependency on the dissipation parameter $\delta$ and therefore can provide a better indication of $\pt$-symmetry breaking through the information back-flow.
\\
Beyond the exceptional point, in the Euclidean metric Hilbert space, distinguishability is no longer periodic; in fact, it is known that distinguishability shows a power law decay with time\, \citep{Kawa17}. As a result, the information back-flow to the subsystem is also reduced and we should see a decrease in the Markovianity of such channels. This is exactly what we see in the figure \ref{blpwithg}.
\\
For the case of the time-dependent metric Hilbert space, in the broken $\pt$-symmetry phase, we note that the distinguishability is still invariant with time. However, in this regime, interpreting the change in distinguishability as the signature of information flow can be problematic. This is because, for time-dependent metrics, the interpretation of the operators as observables needs reconsideration. For example, the Hamiltonian operator beyond the exceptional point is not an observable corresponding to the energy of the system (its eigenvalues are complex), and neither can it be seen as a generator of time translation \citep{ali17}.
\subsection{Coin-position entanglement}\label{sec4-2}
Below the exceptional point, since the Hamiltonian is pseudo-Hermitian, the evolution operator $U=e^{-iH}$ is unitary, in the sense of $U^\# U=\mathbb{1}$. Therefore, the purity of the state $\overline{\rho}$ is preserved below the exceptional point: $tr(\overline{\rho}^2)=1\Rightarrow tr((U\overline{\rho}U^\#)^2)=1$. For such evolutions, therefore, the von Neumann entropy of the reduced density matrix is a measure of the bipartite entanglement, called the entanglement entropy, between the coin and position space.
\begin{figure}
\begin{subfigure}{.5\textwidth}
	\includegraphics[scale=.35]{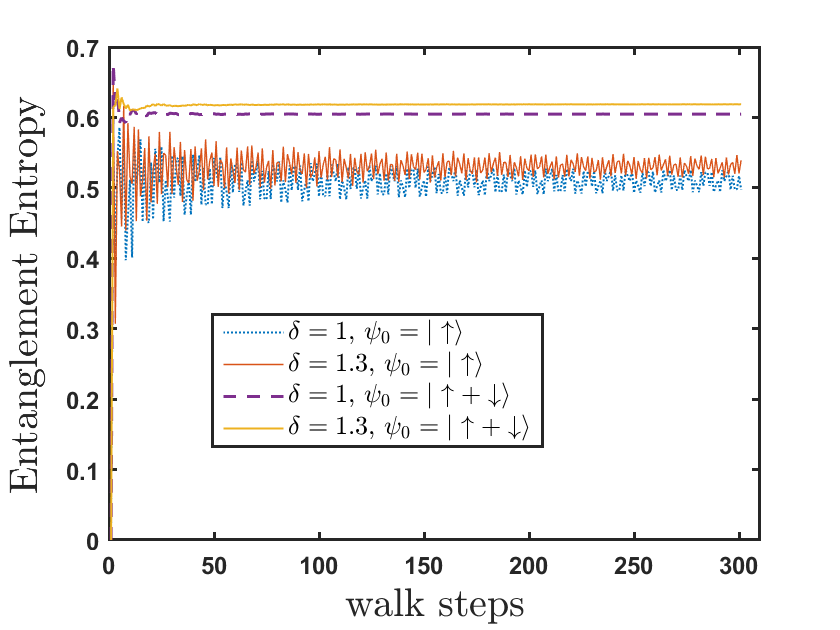}
	\end{subfigure}
\caption{The figure shows the entanglement dynamics of a state as it evolves under the pseudo-Hermitian Hamiltonian. We take two different initial states at the initial position $x=0$, $\psi_0=\ket{\uparrow}$ evolves asymmetrically in the lattice while $\psi_0=\ket{\uparrow}+\ket{\downarrow}$ (written up to normalization) evolves symmetrically. Compared to the Hermitian case, the coin-position entanglement for both walks increases under the pseudo-Hermitian evolution.}
\label{ent}
\end{figure}
The diagonalizable reduced system has a simple form in terms of the eigenvalues of the density matrix. Entanglement entropy is defined as 	$EE(t)=-\sum_i\lambda_i(t) \ln\lambda_i(t)$ where $\lambda_i(t)$ are the eigenvalues of the coin state $\overline{\rho}_c(t)$. Figure \ref{ent} shows the entanglement entropy between the coin and position space for different coin parameters and two different initial states $\ket{\psi_0}$ which correspond to symmetric  $(\ket{\psi_0}=\ket{\uparrow+\downarrow}/\sqrt{2})$ and asymmetric $(\ket{\psi_0}=\ket{\uparrow})$ spread of the probability distribution in the position space. We evolve these states in the metric formalism under a pseudo-Hermitian Hamiltonian and compare it with a Hermitian evolution. We see that for both the choices of initial states, the coin-position entanglement in the large time limit is higher for evolution under the non-Hermitian Hamiltonian. This is a surprising result, since, as discussed in the beginning sections, non-Hermiticity is commonly associated with open system dynamics where the noise from the environment decreases the correlations within the system. An increase in the correlations is possible because the pseudo-Hermiticitian dynamics is equivalent to closed system dynamics. This also hints that pseudo-hermiticity can be seen as a resource in quantum information.
\section{Conclusions}\label{sec5}
In this work, we have studied the interaction between the subsystems of a quantum walk evolving under a $\pt$-symmetric non-Hermitian Hamiltonian. To do this in the metric formalism, we had to give a consistent definition of a subsystem defined through a Hermiticity and norm-preserving map between two Hilbert spaces. The construction of the subsystem uses purely geometric arguments. We show that the subsystem thus constructed is the equivalent to the previous such constructions which were more operationally motivated. The information back-flow into the coin subsystem under both the mechanisms shows a clear power law behavior near the exceptional point, however, the metric method gives a more reliable method of detecting the exceptional point. For the normalization method, the information back-flow to the subsystem can be consistently explained by the information back-flow to the system as a whole. In the end, we also explore the entanglement dynamics between the coin and position subspaces in the metric formalism. Surprisingly, the entanglement between the subsystems increases in the non-Hermitian regime, indicating that pseudo-Hermiticity can be seen as a quantum resource.
\section*{Acknowledgment}
HB thanks Sibasish Ghosh for stimulating conversations on metric formalism and S R Hassan for comments on this work. CMC and SB acknowledge the support from Interdisciplinary Cyber Physical
Systems
(ICPS) programme of the Department of Science and Technology, India, Grants
No.: DST/ICPS/QuST/Theme-1/2019/1 and DST/ICPS/QuST/Theme-1/2019/6,
respectively.

\end{document}